# Development of a Detector Control System for the ATLAS Pixel Detector

S. Kersten, M. Imhäuser, P. Kind, University of Wuppertal, Germany;
H. Burckhart, B. Hallgren, CERN, Geneva, Switzerland;
G. Hallewell, Centre de Physique des Particules de Marseille, France;
V. Vacek, Czech Technical University, Prague, Czech Republic

## Abstract

The innermost part of the ATLAS experiment will be a pixel detector containing around 1750 individual detector modules. A detector control system (DCS) is required to handle thousands of I/O channels with varying characteristics. The main building blocks of the pixel DCS are the cooling system, the power supplies and the thermal interlock system, responsible for the ultimate safety of the pixel sensors. The ATLAS Embedded Local Monitor Board (ELMB), a multi purpose front end I/O system with a CAN interface, is foreseen for several monitoring and control tasks. The Supervisory, Control And Data Acquisition (SCADA) system will use PVSS[1], a commercial software product chosen for the CERN LHC experiments. We report on the status of the different building blocks of the ATLAS pixel DCS.

## 1 OVERVIEW OF THE PIXEL DETECTOR CONTROL SYSTEM

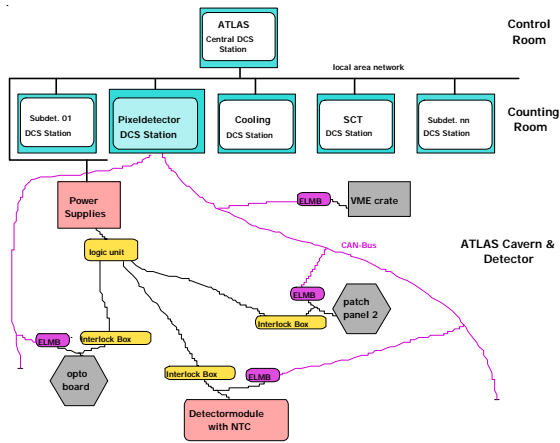

Figure 1: Overview on the pixel detector control system

The ATLAS DCS is organized in 3 levels. The highest handles the overall operation, which is run from the control room on the surface. The intermediate level supervises the sub-detectors and will be run from the underground electronics rooms. These upper two levels of the ATLAS DCS will be implemented in the commercial SCADA system PVSS-II. It comprises the Human Interface and has a modular architecture based on functional units called managers, which perform the tasks of data collection, analysis and trending, the archiving of monitor data, alert handling and "automatic" execution of pre-defined procedures and corrective actions. It is a device-oriented product where devices are modeled by structures called data-points. PVSS-II can exchange data with systems outside of the DCS and has a web interface. Applications can be distributed over many stations on a network running either Linux or WNT/2000. This distribution facilitates the mapping of the control system onto the different sub-detectors.

The lowest level of the ATLAS DCS consists of the front end I/O devices and their field-buses, located in the experimental cavern. Fig. 1 illustrates all three levels and shows the main components of the ATLAS pixel DCS.

The high power density of the pixel front end electronics (up to 0.7W/cm², and ~20KW total heat dissipation), requires a powerful cooling system. An evaporative fluorocarbon-based system has been chosen. The harsh radiation environment puts another constraint on the design of the control system. As heat-ups of irradiated silicon detectors can cause irreparable damage to the sensor material, a thermal interlock ("I-Box") system is implemented, acting on the power supplies and all boards where active components like regulators will be located. I-Box signals are combined and routed by a "logic unit". Around 250 ELMBs will be required to monitor the parameters of the pixel detector, delivering their information to the pixel Local Control Station via the CAN field-bus.

## 2 THE ELMB

A general-purpose, low-cost I/O concentrator, the ELMB has been developed for ATLAS. It is a plug-on board (size 50x67mm) intended for a wide range of different front-end control and monitoring tasks. It is based on the CAN serial field-bus system, is radiation tolerant and can be used in magnetic fields.

The ELMB can either be directly mounted onto sub-detector front-end electronics, or onto a general-purpose motherboard which adapts the I/O signals. The ELMB has an on-board CAN-interface and is in-system programmable, either via an on-board connector or via CAN. There are 18 general-purpose I/O lines, 8 digital inputs and 8 digital outputs. Optionally a 16-bit ADC and multiplexing for 64 analogue inputs is provided on-board. An add-on card containing 16 channels of 12 bit DAC [1] is under development. The ELMB is divided into three sections: analog, digital and CAN. They are separated with opto-couplers to prevent current loops.

---

[1] Prozess Visualisierungs- u. SteuerungsSystem: ETM Co., Wien, Austria

The local intelligence of the ELMB is provided by 2 microcontrollers of the AVR family of 8-bit processors, manufactured by ATMEL. The ELMB's main processor is the ATmega103L running at 4 MHz. The main monitoring and control applications run on this processor. The second on-board microcontroller is a much smaller member of the same AVR family. This processor handles the In-System-Programming (ISP) via CAN of the Atmega103L processor. In addition, it monitors the operation of the ATmega103L and takes control of the ELMB if necessary. This feature is one of the protections against Single Event Effects (SEE).

CAN is one of the three CERN-recommended field-buses [2]. It is especially suited for sensor readout and control functions in the implementation of a distributed control system due to its reliability (in ATLAS, bus lengths of up to 100 m are expected). Further advantages are availability of inexpensive controller chips from different suppliers, ease of use and wide acceptance by industry. The error checking mechanism of CAN is of particular interest in the LHC environment where bit errors due to SEE will occur. CANopen [3] is chosen as the higher layer protocol to standardise data structure and communication. A general purpose CANopen-embedded software program (ELMBio[4]), conforming to the CANopen DS-401 Device Profile for I/O-modules has been developed for the ELMB Master processor.

## 3 THERMAL INTERLOCK SYSTEM

Each pixel detector module will be protected against risks associated with latch-up, de-lamination of a particular module from its cooling channel or failure of coolant flow in the parallel cooling circuit. Each detector module is equipped with its own temperature sensor. Our requirements are best met by a 10 kΩ NTC resistor. The signal from each temperature sensor will be split between an ELMB ADC channel (for data logging) and an I-Box channel, where it is compared to a reference value. In the event of an excessive deviation, a hardwired logic signal acts on the module's power supply (figure 1). In addition to module over-temperature, other error conditions including a broken cable or temperature sensor and a short circuit must also cause the setting of the interlock signal. Negative TTL- compatible 2 bit logic is employed. Compatibility with the ELMB was another design issue.

Several studies [5] have demonstrated that the electrical performance of the circuit is in good agreement with the expected absolute precision, which is composed of the error of the interlock circuit of $\pm 0.2$ K and of the tolerances of the NTC of $\pm 0.3$ K (at 0 to 10 °C).

As the location of the I-Box will be the ATLAS experimental cavern, the radiation tolerance of all its components had to be investigated. Three types of possible problems were studied: damage due to ionising ("total ionising dose": TID) and non ionising radiation ("non-ionising energy loss": NIEL) and single event effects (SEE). Table 1 summarizes the qualification levels (including several safety factors), to which the components had to be tested.

Table 1: "Radiation Tolerance Criteria" for the I-Box

|  | 10 years operation in ATLAS |
|---|---|
| **RTC$_{TID}$** | 93 Gy |
| **RTC$_{NIEL}$** | 4.8 $10^{11}$ n/cm$^2$ (1 MeV) |
| **RTC$_{SEE}$** | 9.22 $10^{10}$ h/cm$^2$ (> 20 MeV) |

The radiation effects were studied with a $^{60}$Co source, 1 MeV neutrons and a 60 MeV proton beam. Various families of NOR-gates and Op-Amps were investigated in order to find the most reliable components. A full set of compatible components for the building of the I-Box is now established. Further details of the study are given in [5].

## 4 THE COOLING SYSTEM

The pixel detector will be cooled by the evaporation of per-fluoro-n-propane ($C_3F_8$). An evaporation temperature of ~-20°C in the on-detector cooling channels will allow a silicon operating temperature of ~ -6°C.

An extensive series of cooling tests in prototype structures has been reported [6]. Following tests with a prototype circulator [7], we have commissioned a large scale (6 kW) demonstrator (Figure 2), supplying up to 25 parallel cooling circuits. Coolant is circulated through interconnecting tubing replicating the lengths and hydrostatic head differences expected in the final ATLAS installation.

The compressor operates between an aspiration and output pressures of ~1 bar$_{abs}$ and ~10 bar$_{abs}$. Aspiration pressure is regulated via PID variation of the compressor motor speed from zero to 100%, based on the sensed pressure in an input buffer tank. $C_3F_8$ vapor is condensed at 10 bar$_{abs}$ and passed to the detector loads in liquid form. A detailed description of the principle of operation is given in ref [7].

Liquid and vapor will circulate via manifolds on the ATLAS service platforms. High radiation levels and magnetic fields predicate that local regulation devices be pneumatic. In each of the ~100 pixel cooling circuits, coolant flow rate will be set from a DAC through a "dome-loaded" pressure regulator, placed ~25m upstream of an injection capillary, piloted by analog compressed air in the range 1-10 $bar_{abs}$ from an "I2P" (4-20mA input) electro-pneumatic actuator. Actuators will receive analog set points from DAC add-ons to the ATLAS ELMB, (section 2). Circuit boiling pressure (hence operating temperature: at 1.9$bar_{abs}$ $C_3F_8$ evaporates at –20°C) will be controlled by a similarly piloted dome-loaded backpressure regulator.

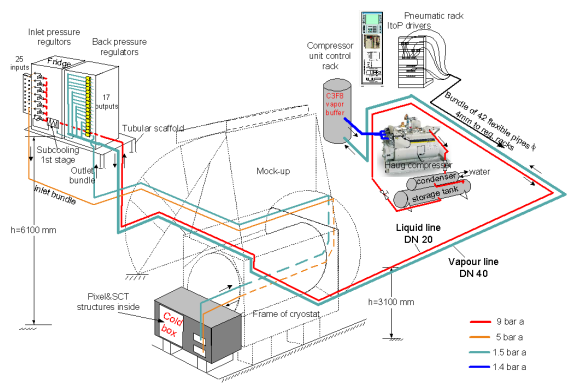

Figure 2: Schematic of the 6 kW Demonstration Evaporative Cooling Recirculator

In the final system, coolant flow in each circuit will be PID-regulated proportional to the variable circuit load. DAC outputs from ELMBs will be used in hardwired feedback loops, the PID algorithm being stored in the onboard ELMB microcontroller chip.

## 5   THE POWER SUPPLY SYSTEM

Seven voltages are necessary for the operation of each pixel detector module and its related optical link. The requirements vary from the depletion bias of up to 700V to several low voltages, some with low power consumption, others drawing 2-5 A.

So-called multi-voltage "complex channels", supplying all voltages necessary for a detector element, are planned. This simplifies the design of the thermal interlock system (section 3): a single interlock signal will simultaneously turn off all power to a module.

We aim to have a power supply system with a high level of local intelligence. A system able to make decisions autonomously - not relying on the functionality of a network - will reduce field-bus traffic and will enhance the safety of the detector. Error conditions (including over-current) will be handled in the power supply system itself.

## 7   SUMMARY

As part of the ATLAS DCS, the pixel control system aims to be an hierarchically coherent system. The SCADA system, based on PVSS, performs the supervisory control, while various front-end monitoring and control tasks will be performed by the ELMBs. An additional level, the thermal interlock system - a pure hardware based solution - is responsible for the safety of the detector.

The necessary hardware components to build the pixel detector control system have been identified, tested and found to fulfil the requirements. The use of the ELMB will further help to reduce the cost for a system containing thousands of channels.

Beam tests will enable us to improve our concept, increase its performance and enlarge the number of control channels toward that of the final system.